\documentclass[12pt]{article}
\title{Dynamics of confined gluons}
\author{ Yu.A.Simonov\\
State Research Center\\Institute of Theoretical and Experimental
Physics, \\ Moscow, 117218 Russia}
\date{}
\newcommand{\beq}{\begin{eqnarray}}
 \newcommand{\eeq}{\end{eqnarray}}
\newcommand{\be}{\begin{equation}}
 \newcommand{\ee}{\end{equation}}

 \def\la{\mathrel{\mathpalette\fun <}}
\def\ga{\mathrel{\mathpalette\fun >}}
\def\fun#1#2{\lower3.6pt\vbox{\baselineskip0pt\lineskip.9pt
\ialign{$\mathsurround=0pt#1\hfil ##\hfil$\crcr#2\crcr\sim\crcr}}}

\newcommand{{\SD}}{\rm SD}
\newcommand{{\GeV}}{\rm GeV}
\newcommand{{\fm}}{\rm fm}
\newcommand{{\oMS}}{\overline{MS}}

\newcommand{\vex}{\mbox{\boldmath${\rm x}$}}
\newcommand{\vew}{\mbox{\boldmath${\rm w}$}}
\newcommand{\vey}{\mbox{\boldmath${\rm y}$}}
\newcommand{\ver}{\mbox{\boldmath${\rm r}$}}

\newcommand{\vep}{\mbox{\boldmath${\rm p}$}}

\newcommand{\vez}{\mbox{\boldmath${\rm z}$}}

\newcommand{\veR}{\mbox{\boldmath${\rm R}$}}

\newcommand{\llan}{\langle\langle}
\newcommand{\rran}{\rangle\rangle}
\newcommand{\lan}{\langle}
\newcommand{\ran}{\rangle}

\begin{document}
\maketitle

\begin{abstract}

Propagation of gluons in  the confining vacuum is studied in the
framework of the background perturbation theory, where
nonperturbative background contains confining correlators.

Two settings of the problem are considered. In the first the
confined gluon is evolving in time together with static quark and
antiquark  forming the one-gluon static hybrid. The  hybrid
spectrum is calculated in terms of string tension and is in
agreement with earlier analytic and lattice calculations. In the
second setting the confined gluon is exchanged between quarks and
the gluon Green's function is calculated, giving rise to the
Coulomb potential modified at large  distances. The resulting
screening radius of 0.5 fm presents a serious problem when
confronting with lattice and experimental data. A possible
solution of this discrepancy is discussed.

\end{abstract}

\section{Introduction}

Gluons are known to be confined but this property is never taken
into account in the Standard Perturbation Theory (SPT). As an
argument one refers to the small distance (high momentum) domain
where SPT is assumed to be valid. Beyond this domain the SPT
displays unphysical singularities, and moreover the  very notion
of gluon should be properly defined. This can be  done in the
framework of the  Background Perurbation  theory (BPT) \cite{1}.
The formalism of this kind where the background is assumed to be
nonperturbative with confining properties, was developed in
\cite{2,3}.

One immediate consequence of this new BPT is that all gluons are
confined and moreover the unphysical singularities of SPT (Landau
ghost poles and branch points as well as IR renormalons) are
removed from the theory \cite{2,3}.

 The confined gluons can form several types of systems: glueballs
 \cite{4}, hybrids \cite{5,6}
and gluelumps \cite{7}. The analytic calculations in the quoted
papers predict spectrum which in all cases is in good agreement
with lattice data, and has a very simple form depending only on
 string  tension     $\sigma $ and $\alpha_s$.

 The main subject of the  present paper is the study of the gluon
 exchange interaction  between $Q$ and $\bar Q$ when gluon is
 confined, which can be called the confined Coulomb interaction
 $V^*_c(R)$. One can expect that at small distances,  say $R<1$
 GeV$^{-1}$, the confined Coulomb potential coincides with the
 standard Coulomb potential $V_c(R) =-C_2\frac{\alpha_s(R)}{R},~~
 C_2=\frac{N^2_c-1}{2N_c}$.

At large $R$ the  confined gluon is expected to produce the
screening of the Coulomb interaction. At the first glance the
screening mass should coincide with the lowest hybrid  mass  of
the static hybrid. However as will be shown below, this naive
expectation fails, since gluon propagation between quarks goes not
only in time (where hybrid mass is in proper place) but also in
distance $R$ (where in addition asymptotics of wave functions
enters). As a result one obtains a more complicated behaviour
which we display both numerically and analytically.

The plan of the paper is as follows. In section 2 the general
formulas of BPT are written down and approximations are discussed.
In section 3 a simple toy model is suggested to illustrate method
and possible qualitative outcome.

In section 4 the method is applied to calculate the static hybrid
Green function and spectrum in a way different from used before,
in \cite{6}. In section 5 results of previous section are used to
calculate the Green function of the  confined gluon exchanged
between static quarks, and the  resulting screened Coulomb
potential. Physical consequences and prospectives are discussed in
the concluding section.

\section{Dynamics of a confined gluon. Generalities.}

In the Field Correlaror Method  (FCM) the dynamical picture of a
confined gluon is simple and selfconsistent: the gluon (its
corresponding field is $a_\mu$) moves in the  strong and
disordered vacuum field $B_\mu$ characterized by the  correkators
of the field  $F_{\mu\nu} (B)$, so that the total gluonic field
$A_\mu$ can be written as \be A_\mu=B_\mu+a_\mu. \label{D1}\ee
Here the problem of separation of
 $A_\mu$ into $B_\mu$ and $a_\mu$, and that of double counting is
 resolved technically by the use of the so-called 'tHooft identity
\cite{3} and one can integrate and average over both $DB_\mu$ and
 $Da_\mu$, so that the partition function is
 \be
 Z=\frac{1}{N'} \int DB_\mu Z(J,B),~~ Z(J,B) =\int Da_\mu \exp
 (-S(a+B) +J(a+B))\label{D2}\ee and $S$ is the standard QCD
 Euclidean action including ghost  and gauge-fixing terms (see
 \cite{2,3}
 for details).

 In what follows we shall be interested in the gluon propagation
 in the field of static (confined) quark $Q$ and antiquark $\bar
 Q$. The starting point for the gauge-invariant study of this
 process is the total Green's function of $Q\bar Q $ system, which
 is proportional to the Wilson loop:
 \be \lan W(A)\ran \equiv \llan W(B+a)\rran_{B,a} =\llan W(B+a)\ran_ a \ran _B,\label{D3}\ee
\be W(A) =tr ~P\exp ig \int_CA_\mu dz_\mu.\label{D4}\ee

The Wilson loop is assumed to be the closed rectangular contour
$R\times T_0$ in the ($x_1, x_4$) plane.

One may expand in $ga_\mu$ keeping  $B_\mu$ intact and this will
give the Background Perturbation Theory (BPT) series (\cite{1,2},
see \cite{3} for details), the first terms being \be \lan
W(B+a)\ran_{B,a} = \lan W^{(0)} (B)\ran_B - g^2 \lan W^{(2)}(B,
x,y) G_{\mu\nu} (x,y) \ran_B d x_\mu dy_\nu+...\label{D5}\ee For
the chosen contour C and the  Feynman gauge of field $B_\mu$ and
FFSR for $G_{\mu\nu}, W^{(2)}$ and $G_{\mu\nu}$  can be written as
\be W^{(2)} = C_2 (f) \Phi^{(-)}_{\alpha\beta} (x,y)
\Phi^{(+)}_{\alpha'\beta'} (y,x)\label{D6}\ee \be G_{\mu\nu} (x,y)
=\int^\infty_0 ds (Dz)_{xy} e^{-K}\left\{\Phi^{(adj)}(x,y)  P\exp
(2g \int^s_0 F_{\rho\sigma} (z(\tau))
dz)\right\}^{(\mu\nu)}_{\beta\alpha, \beta'\alpha'} \label{D7}\ee
where $\Phi^{(\pm)} (x,y)$ are the future/past pieces of the
Wilson loop $W^{(0)}(B)= P\exp (ig \oint B_\mu dz_\mu)$, obtained
by cutting it at points $x$ and $y$. Note that  the adjoint phase
factor $\Phi^{(adj)}{(x,y)} \equiv \exp (ig \int^x_y B_\mu
dz_\mu)$ from $G_{\mu\nu}$ which can be written as the product
\be
 \Phi^{(adj)} (x,y) = \Phi_{\beta\alpha} (y, x)
 \Phi_{\beta'\alpha'} (x,y)\label{D8}\ee
 the total construction produces two closed Wilson loops (see ref.
 \cite{2,3,6} for pictures and discussion). \be
  \Phi^{(-)} \Phi^{(adj)} \Phi^{(+)}=W^{(-)}_\sigma (x,y)
  W^{(+)}_\sigma (y,x).
  \label{D9}\ee
  Note the subscript $\sigma$ in (\ref{D9}) which implies the
  color-magnetic spin factor (the last factor on the r.h.s. of (\ref{D7})) entering in all Wilson lines
  including $\Phi^{(adj)}$. The averaging of (\ref{D9}) over
  fields $B_\mu$ can be easily     done  at large  $N_c$
  \be \lan W^{(2)}_{  \mu\nu}G_{\mu\nu}\ran_B =   G^{(0)}\lan
  W^{(-)}_\sigma (x,y) \ran_B \lan W_\sigma^{(+)} (y,x)\ran_B
, \label{D10}\ee where $G^{(0)}$ denotes the integral $\int ds
(DZ)_{xy} e^{-K}$.
  In the FCM one obtains for $\lan W \ran$ the area law behaviour
  at large distances $R, T\gg \lambda$\be\lan W\ran =\exp
  (-\sigma  S_{min})\label{D11}\ee
 and $\lambda$ is the gluon correlation length \cite{8},
  characterizing the fall-off in $x$ of the field correlator $D(x)
  \sim \lan tr F(x) \Phi F(0)\ran $, and $S_{min}$ is the area,
  which is assumed to be minimal for the given contour $C$.

Note that at large distances the spin factor does not contribute
to the area law [4-7] and therefore the subscript $\sigma$ in
(\ref{D11}) is omitted.

   To describe $S_{min}$ one can parametrize first the trajectory
   $z_\mu(t)$ of the gluon between the  initial point $x$ and the
   final point $y$ (both on the contour $C$).
   \be z_\mu:( z_1 (t) \equiv \xi (t),~~ z_2(t), ~~ z_3(t), z_4(t)\equiv
   t),~~ h^2(t) = z^2_2 (t) +z^2_3 (t)\label{D12}\ee
   We choose the  Nambu-Goto ansatz for the minimal area surface (or
   rather for its increase over the plane area $S_0=RT$) \be
    \Delta S = \Delta S_1+\Delta S_2,~~ \Delta S_i= \int^{x_4}_{y_4} dt \int^1_0 d\beta  \sqrt{(\dot{ w}_i
    w'_i)^2-\dot{w}_i^2 w^{'2}_i}
    \label{D13}\ee
where $w_{i\mu} (\tau,\beta): w_{i4}=t, \vew_{1,2}
=(1-\beta)(\veR,0)+\beta \vez$.
   Note, that in our case two different processes can be initiated
  by the exchanged gluon:

  i) points $x$ and $y$ are at $x_4=T_0, \vex=(\frac{R}{2},0,0)$ and $y_4=0,~ \vey=(\frac{R}{2},0,0)$, the
  situation which is insured  on the lattice by the insertion of
  plaquettes at these sides of Wilson loop. In this case one
  obtains the  hybrid excitation of the Wilson loop, and this form was
  used before to compute hybrid spectra analytically \cite{5,6} and  on
  the lattice \cite{9}.

  ii) the points $x$ and $y$ belong to the trajectories of $Q$ and
  $\bar Q$ respectively, so that $G_{\mu\nu}(x,y) = G_{44} (R,T),
  ~~ T< T_0$, describes the propagation of Coulomb gluon
  between the quarks.

  In what follows we shall study both cases using for that the
  final form resulting from (\ref{D5}), (\ref{D7}), (\ref{D11}),
  (\ref{D13})
  \be \lan G_{44} (x,y)\ran_B \equiv G(x,y) = \int^\infty_0 ds (Dz)_{xy} e^{-K-\sigma \Delta
  S}\label{D14}\ee

  To treat the awkward roots in $\Delta S,$ Eq. (\ref{D13}), one
  can introduce the einbein parameters $\nu(t)$ and $\bar \nu(t)$
\cite{10} obtaining  in the small oscillation limit
  \be
   G(x,y) =\int^\infty_0 ds (Dz)_{xy} D\nu D\bar \nu
 e^{-K-\sigma \Delta \tilde S(\nu,\bar \nu)}\label{D15}\ee
 where \be \Delta \tilde S (\nu,\bar \nu)=\int^{x_4}_{y_4} dt \left\{
 \frac{\nu+\bar \nu}{2} \left(1-\frac13\dot{z}^2_\bot\right)+\frac{h^2+\xi^2}{2\nu} +
 \frac{h^2+(R-\xi)^2}{2\bar \nu} -R\right\} \label{D16}\ee

For $x=\left( \frac{R}{2}, 0,0, T_0\right ); ~~y=\left(
\frac{R}{2}, 0,0,0\right) $ one will have from (\ref{D14}) the
static hybrid spectrum, to be compared with previous calculations,
and for the case  ii) one can define the generalized Coulomb
interaction as
\be
\int^{T_0}_0 dx_4 \int^{T_0}_0 dy_4 G(x,y) = \int^{T_0}_0
d\frac{x_4+y_4}{2}\int^{T_0/2}_{-T_0/2} G (R,T) dT\equiv T_0
V_C^*(R).\label{D17}\ee

In the limit $\sigma \to 0$ one obtains the free gluon propagator
$G^{(0)}(x, y) =\frac{1}{4\pi^2 (x-y)^2}$ in case  i) and the
Coulomb interaction $V_c^* (R) = V_c(R) = -C_2 \frac{\alpha_s}{R}$
in the case  ii). Our purpose in what follows is to obtain
modification of these results for nonzero $\sigma$. To grasp the
idea we shall start with a toy model as a warm-up.

\section{A simple toy model for the confined gluon}

Consider a gluon propagating from static quark $Q$ with coordinate
$(0, \vec 0)$ to antiquark $\bar Q (T, R, 0,0)$.

The world sheet of $Q\bar Q$ system with the  string from $Q$ to
$\bar Q$ sweeps the strip in  the ($x_4, x_1)$ plane, and one
expects that  gluons are confined dynamically to some region
around this strip.
 This means that the running away of gluons from the plane ($x_4,
 x_1)$ is damped by some function, which we take in the form of
 the confining "potential" $V$,
 \be
  V(\vez) = \frac{\omega^2}{4} (z^2_2+z^2_3).\label{D18}\ee
   The Green's function of the gluon can be written in the FFSR
   \cite{11}
   \be
   G(T,R) =\int^\infty_0 ds (Dz_1)_{0R}(Dz_4)_{0T} (Dz_2)_{00}
   (Dz_3)_{00} e^{-K-\int^s_0 V d\tau}\label{D19}\ee
   where $K=\frac14 \int^s_0 d\tau \left(
   \frac{dz_\mu}{d\tau}\right)^2$, and $(Dz_i)_{ab} =\prod^N_{n=1}
   \frac{dp_i}{2\pi} e^{ip_i (a_i-b_i)} \left(\frac{d\Delta
   z_i(n)}{4\pi \varepsilon}\right)$.

   The integration in $Dz_i$ factorizes and can be performed
   immediately, with the result (see Appendix 1 for details)
   \be G(T,R) =
   \frac{1}{16\pi^2} \int^\infty_0 \frac{ds}{s^2} \varphi (\omega s) \exp
   (-\frac{x^2}{4s}),~~ \varphi(t) =\frac{t}{\sinh t}; ~ x^2 =R^2 +
   T^2.\label{D20}\ee

The integral (\ref{D20}) can be estimated by the stationary point
method  and one obtains
\be
G(T,R) \cong \frac{\psi(\omega  x^2)}{4\pi^2 x^2}, \psi (t)
\approx \left\{ \begin{array}{ll}\frac{t}{8\sinh\left(
\frac{t}{8}\right)} , & t\ll 1\\ \sqrt{t}e^{-\sqrt{t}},& t\gg
1\end{array}\right..\label{D21}\ee

One can see that at large $x^2$, $x^2 \omega \gg 1$, there appears
in (\ref{D21}) the damping factor signalizing the mass gap
$m=\sqrt{\omega}$ , i.e. the confined gluon acts at large
distances as a massive particle, while at small distances, $x^2
\to 0 $, it behaves as the ordinary unconfined gluon. When
$\omega\to 0$ one recovers from (\ref{D17}) the standard Coulomb
interaction: $$V_c(R) =- g^2 C_2(f) 2 \int^\infty_0 G(T,R) dT =-
\frac{\alpha_s C_2(f)}{R},$$ and in the opposite  limit,  $\omega
R^2\gg 1$, $V_c(R)$ is multiplied by the factor
$\frac{4\sqrt{2}}{(\omega R^2)^{1/4}}\exp (-(\omega R^2)^{1/2})$.
This screening factor is equal $1/2$ at $R\approx
\sqrt{\frac{4}{\omega}}$ and decays  exponentially at large $R$,
as it is commonly expected. In section 5 a more complicated
behaviour will be obtained in the realistic case when gluon is
confined by the string world-sheet. To this end we develop in the
next section the necessary formalism for the hybrid Green
function.

\section{Spectrum of the confined gluons. Static hybrid}

In this section we shall calculate the gluon Green's function in
the hybrid situation, i.e. when boundary conditions are given in
i) below Eq. (\ref{D13}), and one can identify $T\equiv T_0$. One
can introduce the einbein variable $\mu(t)$ as in \cite{10} (see
Appendix 2 for details) and write
\be
G(x,y) =\int \frac{D\mu}{2\bar \mu} D\nu D\bar \nu e^{-\Gamma} G_3
(R,T,\nu,\bar \nu, \mu)\label{D3.1}\ee where we have defined
\be
\Gamma=\int^T_0 \frac{\mu}{2} dt +\frac{\sigma}{2} \int^T_0
(\nu+\bar\nu) dt +\sigma R^2 \int^T_0 \frac{dt}{2(\nu+\bar
\nu)},\label{D3.2}\ee \be G_3=\int(D^3z)_{xy} e^{-\int^T_0
\left(\frac{\mu}{2} {\dot
{z}}_1^2+\frac{\mu+J_1+J_2}{2}\dot{z}^2_\bot\right)  dt
-\frac{\sigma}{2\tilde \nu} \int^T_0 [( z_1-\tilde R)^2+ h^2]
dt},\label{D3.3}\ee $\tilde R=R\frac{\nu}{\nu+\bar \nu}$, and we
take $x=\left(\frac{R}{2}, 0,0,0\right), ~~ y=\left(\frac{R}{2},
0,0,T\right),~J_i=\frac{\sigma}{3}\nu_i$.

The integration over $(D^3z)$ can be immediately done using the
standard path integral formula, see Appendix 1. $$ G_3= \left(
\frac{\mu\omega_1}{2\pi \sinh\omega_1 T}
\right)^{1/2}\frac{(\mu+J_1+J_2)\omega_\bot}{2\pi
\sinh(\omega_\bot T)}$$ \be \exp \left\{ -
\frac{\mu\omega_1}{2\sinh\omega_1 T} \frac{R^2(\nu-\bar
\nu)^2}{2(\nu+\bar\nu)^2} (\cosh (\omega_1
T)-1)\right\}\label{D3.4}\ee where
$\omega_1=\sqrt{\frac{\sigma}{\mu\tilde \nu}},~~ \omega_\bot=
\sqrt{\frac{\sigma}{\tilde\nu(\mu+J_1+J_2)}}, ~~ \tilde \nu=
\frac{\nu\bar \nu}{\nu+\bar\nu}$.

The next step is the stationary point analysis of Eqs.
(\ref{D3.1}), (\ref{D3.4}) with respect to variables $\nu,\bar
\nu, \mu$. We relegate the details of this analysis to the
Appendix 2 and here only quote the result. Minimizing  the
expression in the exponent at (\ref{D3.4}) with respect to
$(\nu-\bar \nu)$ one obtains that  $\nu=\bar\nu$ , and the
stationary point in $\nu$, $\nu=\nu_0$ is found from the equation
\be
\frac{\partial}{\partial\nu}(\sigma\nu +\frac{\sigma
R^2}{4\nu}+\frac{\omega_\bot}{2}+{\omega_\bot)=0 }.\label{D3.5}\ee
One can distinguish two cases: a) $\sigma R^2\gg 1$; b) $\sigma
R^2\ll 1.$

a) In the case a) taking into account that $\mu_0\la
\sqrt{\sigma}$ (which will be confirmed afterwards), one has
\be
\nu_0 =\frac{R}{2},~~ \omega^{(0)}_1\to \left( \frac{4\sigma}{\mu
R}\right)^{1/2},~~ \omega_\bot^{(0)}\to
\frac{\sqrt{12}}{R}\label{D3.6}\ee and Eq.(\ref{D3.1}) with
$\nu=\bar\nu=\nu_0$ assumes the form for large $\omega T$
\be
G(x,y)=\int D\mu \exp [-(\sigma R+ \omega_\bot^{(0)}+\frac{\mu}{2}
+ \frac{1}{2} \sqrt{\frac{4\sigma}{\mu R}} )T]\label{D3.7}\ee The
stationary point $\mu=\mu_0$ is found from the exponential of
(\ref{D3.7}) to be $$ \mu_0 = \left(\frac{\sigma}{R}\right)^{1/3},
$$ and the resulting static hybrid mass at large $R$ is
\be
M_{hybrid} (R) =\sigma R+ \frac32
\left(\frac{\sigma}{R}\right)^{1/3}+
\frac{\sqrt{12}}{R}\label{D3.8}\ee

 The second term on the r.h.s. of (\ref{D3.8}) is the characteristic "$R^{-1/3}$ law" for large $R$ hybrid
excitations,  studied in \cite{6}, and one can distinguish the
longitudinal and transverse branches of spectrum with higher
excitations generated by $\sinh\omega_1 T$ and $\sinh\omega_\bot
T$ in (25), which we write as in \cite{6}
 \be M_{hybrid}^{(long)} = \frac{3}{2^{1/3}}\left(\frac{\sigma}{ R}\right)^{1/3}\left(n_z+\frac12\right)^{2/3},~~
  M^{(trans)}_{hybrid} =\frac{\sqrt{12}}{R}(n_\bot+\Lambda+1),
\label{D3.8'}\ee where $\Lambda$ is angular momentum projection on
the $x$ axis. Note that $\sqrt{12}=3.46\approx \pi$ and transverse
spectrum is very close to the flux tube excitations, while
 the longitudinal found in \cite{6} is new.

 The resulting spectrum
(\ref{D3.8'}) is  in   a good agreement with lattice calculations
\cite{9}, see also discussion in
 \cite{6}.

b) We now turn to the case b) $\sigma R^2\ll 1$, and from
(\ref{D3.5}) find that
$\nu=\nu_0=\left(\frac{9}{\sigma\mu}\right)^{1/3}$ and the
equivalent of Eq. (\ref{D3.7}) is
\be
G(x,y) =\int D\mu \exp \left\{ -\left[ \frac{\mu}{2} + \frac32
\frac{(3\sigma)^{2/3}}{\mu^{1/3}}+ \frac{\sigma R^2}{2}\left(
\frac{\sigma\mu}{9}\right)^{1/3}\right] T\right\}.\label{D3.9}\ee
To check accuracy of our approach we can calculate the hybrid mass
in the limit $R\to 0$, which coincides with the gluelump case
\cite{7}. Defining for $R \to 0  $ the stationary point
$\mu=\mu_0$ from the exponent of (\ref{D3.9}) one has
$\mu=\mu_0=\sqrt{3\sigma}$ and the gluelump mass is \be
M_{gluelump}=2\sqrt{3\sigma}\label{D3.10}\ee which should be
compared with  the dedicated gluelump calculations in \cite{7}:
$M_0
=2\left(\frac{a}{3}\right)^{3/4}\left(2\sigma_{adj}\right)^{1/2}=2\sqrt{3.096
\sigma}$, where we have used the value of the first zero of the
Airy function, $a=2.338$. One can see agreement on the level of
1.5\%.

Taking into account the last term in the exponent of (\ref{D3.9})
one obtains the lowest hybrid mass at small $\sigma R^2\ll1$\be
M_{hybrid} (R) =2 \sqrt{3\sigma} + \frac{\sigma R^2}{2}
\sqrt{\frac{\sigma}{3}}+O(R^4)\label{D3.11}\ee which coincides
with the mass spectrum obtained in \cite{6} for this limiting case
by a different method.

Thus our approach can be used as a good zeroth order approximation
for the confined gluon Green's function and its spectrum.

In what follows we shall use the  dependence  $\mu_0(R) =\bar \mu$
given above in two limiting cases:  \be
\mu_0(R)=\left(\frac{\sigma}{R}\right)^{1/3}, ~~ \sigma R^2\gg
1\label{D3.12}\ee $$\mu_0(R)=\sqrt{3\sigma},~~ \sigma R^2\ll 1$$

\section{The confined Coulomb interaction}

In this section we consider the confined gluon Green's function
for the initial and final conditions corresponding to the Coulomb
gluon exchange.

With the same notations as in (\ref{D3.1})-(\ref{D3.3}), one has
\be G(R,T) =\int D\nu D\bar \nu  \frac{D\mu}{2\bar \mu}e^{-\Gamma}
G_3^{(C)} (R,T, \nu, \bar \nu, \mu)\label{D4.1}\ee where $\Gamma$
is the same as in (\ref{D3.2}), but  now $G_3^{(C)}$ is not given
by (\ref{D3.4}), but has another form, due to different initial
and final condition: $x=(0,0,0,0),~~ y=(R,0,0,T)$ and the same
simple general formula of Appendix 1, yields $$
G_3^{(C)}=\left(\frac{\bar \mu\omega_1}{2\pi \sinh \omega_1
T}\right)^{1/2} \frac{(\bar\mu+J_1+J_2)\omega_\bot}{2\pi
\sinh(\omega_\bot T)}$$ \be \exp \left\{ -\frac{\bar
\mu\omega_1}{2\sinh(\omega_1 T)} [R^2 \cosh \omega_1 T+ 2 R^2
\frac{\tilde \nu}{\nu+\bar \nu}(1-\cosh \omega_1 T)]\right\}
\label{D4.2}\ee where we have defined $\omega_\bot$ as in previous
section,  while $$ \omega_1 =\sqrt{\frac{4\sigma}{\bar \mu
(\nu+\bar \nu)}},~~ D\mu = \prod_n\frac{d\mu(n)\sqrt{\Delta
t}}{\sqrt{2\pi\mu(n)}},$$ so that $\int D\mu\exp \left\{-
\frac12\int^T_0 \mu (t) dt\right\}=1$. Minimizing in $(\nu-\bar
\nu)$, one obtains $\nu=\bar\nu, \omega_1\to
\omega_1^{(0)}=\sqrt{\frac{2\sigma}{\bar \mu\nu}}$,
$\omega_\bot\to
\omega_\bot^{(0)}=\sqrt{\frac{2\sigma}{(\bar\mu+J_1+J_2)\nu}}$ and
one has
\be
G(R,T)=\int\frac{D\nu}{(2\pi)^{3/2} 2\bar \mu} \exp
(-\Gamma_0)\label{D4.3}\ee where $$ \Gamma_0 = \sigma\nu T +
\frac{\sigma R^2 T}{4\nu} +\frac 12 \ln \sinh (\omega_1^{(0)} T)
-\frac12 \ln  (\bar \mu \omega_1^{(0)}) +\frac{\bar \mu R^2 }{2T}
\varphi(\omega_1^{(0)} T)+$$ \be+ \ln
\sinh(\omega_\bot^{(0)}T)-\ln \left((\bar
\mu+J_1+J_2)\omega_\bot^{(0)}\right), \label{D4.4}\ee and \be
\varphi (x) = \frac{x(1+ \cosh x)}{2\sinh x}, ~~ \varphi(0)=1,
~~\varphi (x\to \infty) \approx \frac{x}{2}\label{D4.5}\ee

To proceed one should find the stationary point of $\Gamma_0$ with
respect to $\nu$, $\left.\frac{\partial \Gamma_0}{\partial
\nu}\right|_{\nu=\nu_0}=0$. This is easy to do at large $R$, since
then $\nu_0\sim \frac{R}{2}$ and $\omega_1^{(0)}
=\sqrt{\frac{2\sigma}{\bar \mu\nu_0}} \to 0$,
$\omega_\bot^{(0)}\to 0$ so that the last three terms on the
r.h.s. of (\ref{D4.4}) do not contribute. \be \frac{\partial
\Gamma_0}{\partial \nu} =0 =\sigma T \left(1-
\frac{R^2}{4\nu^2_0}\right), ~~ \nu_0 =\frac{R}{2},~~ R\to
\infty\label{D4.6}\ee and $$ \Gamma_0 (\nu=\nu_0) -\sigma
RT=\frac12 \ln \sinh \left(\sqrt{\frac{4\sigma}{\bar \mu R}}
T\right) -\frac12\ln \left(\bar\mu\sqrt{\frac{4\sigma}{\bar \mu
R}}\right)+\frac{\bar \mu R^2}{2T} \varphi
\left(\sqrt{\frac{4\sigma}{\bar \mu R}}T\right)+$$\be  \ln \sinh
\left(\frac{\sqrt{12}}{R} T\right) -\ln \left(\bar \mu
+\frac{\sigma R}{3}) \frac{\sqrt{12}}{R}\right) .\label{D4.7}\ee

To get the modified Coulomb interaction at large $R$, one
considers the integral \be V^*(R) \equiv\int^\infty_{-\infty} dT
G(R, T)= 2\int^\infty_0 dT G (R,T) =\frac{2}{(2\pi)^{3/2} 2\bar
\mu}\int^\infty_0 e^{-\Gamma_0} dT .\label{D4.8}\ee Inserting
(\ref{D4.7}) into (\ref{D4.8}) one obtains $$
V^*(R)=\frac{2\sqrt{3}\left( \frac{\bar
\mu}{R}+\frac{\sigma}{3}\right)}{2\pi^{3/2}}\int^\infty_0
\frac{dT}{\sinh \left(\frac{\sqrt{12}T}{R}\right)\left[\sinh
\left(\sqrt{\frac{4\sigma}{\bar\mu
R}}T\right)\right]^{1/2}}$$\be\exp \left(-\frac{\bar \mu R^2}{2T}
\varphi \left(\sqrt{\frac{4\sigma}{\bar\mu
R}}T\right)\right).\label{D4.9}\ee

Introducing new variable $\tau=\frac{\bar \mu R^2}{2T}$ this
integral can be reduced to the form
\be
V^*(R) =\frac{\sqrt{3} \left( (
{\bar\mu}{R})^2+\frac{\lambda^2}{3}\right)}{2\pi^{3/2}R}
\int^\infty_0 \frac{d\tau}{{\tau^2}}
\frac{e^{-\tau\varphi\left(\frac{\lambda}{\tau}\right)}}{
\left({\sinh\frac{\lambda}{\tau}}\right)^{1/2} \sinh \left(
\frac{\sqrt{3}\bar \mu R}{\tau}\right)}\equiv\frac{\xi (R)}{4\pi
R} \label{D4.10}\ee where we have defined $\lambda= (\sigma \bar
\mu R^3)^{1/2}{\to}_{{R\to \infty}} (\sigma R^2)^{2/3}$. Finally
$\xi(R)$ can  be written as
\be
\xi(R) =\sqrt{\frac{3}{\pi}} 2\lambda \left( 1+\frac{\lambda}{3}
\right)  f(\lambda), ~~\lambda = (\sigma R^2)^{2/3},\ee \be
f(\lambda)= \int^{\infty}_0 dy \frac{e^{-\varphi(\lambda
y)/y}}{\sqrt{\sinh (\lambda y)}\sinh (\sqrt{3\lambda} y)}\ee
 For
$\lambda  \to 0$ one  has  $f(\lambda)\approx \frac{1}{2\lambda}
\sqrt{\frac{\pi}{3}}$, and $\xi (\lambda \to 0)$ is close to
unity.  The explicit behaviour of $\xi(R)$ is given in  the
Table.\\

\begin{center}
\vspace{0.5cm}
\begin{tabular}{lp{11cm}}
{\bf Table }:&  The screening  factor $\xi (R)$ and $f(\lambda)$
as functions of distance $R$
\\

\end{tabular}

\begin{tabular}{|c|c|c|c|c|c|c|c|c|}
\hline
$\lambda$ & 0.1 & 0.2 & 0.4 & 0.6 & 0.8 & 1.0 & 2.7 & 5.4 \\\hline
$f(\lambda)$& 4.60 & 1.91 & 0.74 & 0.41 & 0.26 & 0.18 & 0.024 &
0.0029\\\hline $R$ (fm) & 0.084 & 0.14 & 0.23 & 0.32 & 0.39 & 0.46
& 0.98 & 1.645 \\\hline $\xi(R)$  & 0.929 & 0.796 & 0.656 & 0.577
& 0.515 & 0.469
 & 0.241 &0.086 \\ \hline
\end{tabular}
\end{center}
\vspace{1cm}

 For $\sigma R^2 \to \infty$,
one has from (\ref{D3.12}) $ \bar \mu = \left(
\frac{\sigma}{R}\right)^{1/3}$ and inserting this into (44) one
obtains the asymptotics.
 \be V^*(R) \approx \sqrt{\frac{2}{{3}\pi}}\frac{\lambda \exp
(-\frac{\lambda}{2})}{ R} \approx \frac{\sqrt {2}(\sigma
R^2)^{2/3}\exp\left[-\frac12(\sigma
R^2)^{2/3}\right]}{\sqrt{3}\pi^{3/2} R},~~ (\sigma R^2)\gg 1
\label{D4.11}\ee
 One can see  in the Table that the
 screening starts at rather small  values of $R$ and at  $R\sim 0.5$
 fm, the coefficient $\xi(R) \sim 0.5.$

\section{Conclusions}

The overall static interaction between $Q$ and $\bar Q$ can be
derived from Eq. (\ref{D5}), where the term $\lan W^{(0)} (B)\ran$
gives the confining term $V_{conf}(R)$, while the second term on
the r.h.s. of (\ref{D5}) provides the screened Coulomb potential
$V^*_C(R)$, so that one has for $R \gg \lambda$, $\lambda \approx
0.2$ fm. \be V_{static}(R)=\sigma R+V^*_C(R), V^*_C(R) =-
\frac{C_2\alpha_s}{R} \xi(R),\label{D48}\ee where $\xi(R)$ is
given in (45),(46) and in the Table.

 In (\ref{D48}) the
interference of perturbative field $a_\mu$ and nonperturbative
$B_\mu$ is taken into account. At small $R, R\la \lambda$,  there
exists another interference effect, which was treated before in
\cite{12}, and which provides linear  behaviour of $V_{conf}(R)$
at very small $R$, while without this interference
$V_{conf}(R)\sim const R^2$, for $R\la \lambda $ \cite{13}.

The behaviour $\sigma R-\frac{e}{R}$ was checked on the lattice in
the interval 0.1 fm $<R<$1 fm with good accuracy \cite{14} and the
region  0.8 fm $<R<$1.5 fm was also measured \cite{15} in the
regime where  string breaking is expected. Recently a thorough
analysis of the region $R\la 1 $ fm [16] has revealed that
$\xi(R)$ is approximately constant in this interval and coincides
within 15\% with the bosonic string Casimir
 energy prediction (see [16] for discussion and earlier references). Also the heavy quarkonia spectrum calculated
 assuming $\xi(R)\equiv 1$ is in  good agreement with experiment
 including high excited bottomonium states  [17,18] and deviation of
 $\xi(R)$ from unity for $R\ga 0.5$ fm  seems to  deteriorate this
 agreement. Thus the calculated in this paper screening factor
 $\xi(R)$, Eq.(45) and Table,  is in a serious conflict with
 lattice and phenomenological (experimental) data. A possible
 solution of this paradox lies in adopting the bosonic string term
 [16] at  distances beyond 0.5 fm, so that the sum of the
 screened Coulomb term  and  the bosonic string term  could imitate the original unscreened Coulomb
 interaction. This picture of trasmutation of the Coulomb  into
 the string vibration term, if realistic, can be supported by the
 Casimir scaling study of the Coulomb-like term at distances
 around 1 fm. The accuracy  of the previous study by Bali [14] was
 insufficient
 to draw definite conclusions about the presence of the
 bosonic string term in this region.

 In addition one should recalculate the bosonic string term using
 the realistic hybrid spectrum found in [6], which will be
 reported elsewhere.

 The author is  grateful for useful discussions to A.M.Badalian, N.O.Agasyan,
 A.B.Kaidalov, Yu.S.Kalashnikova and V.I.Shevchenko.

 The work  is supported
  by the Federal Program of the Russian Ministry of industry, Science and Technology No.40.052.1.1.1112.\\

\vspace{2cm}

{\bf Appendix 1}\\

{\bf Derivation of the gluon Green's function in the toy model of
the confining plane (chapter 3) }\\

 \setcounter{equation}{0} \def\theequation{A1.\arabic{equation}}
Eq.(19)  describes Green function of free motion in the plane
($x_1, x_4)$ and factorizable motion in oscillator potential in
the directions $x_2$ and $x_3$. For the latter one can use the
standard textbook formula (see e.g. \cite{16}) for the Green's
function $G(x_a, t_a; x_b, t_b)$ corresponding to the Lagrangian
$$ L=\frac{m\dot{x}^2}{2} - \frac{m\omega^2_0 x^2}{2},$$ which we
write  in the  Euclidean metrics $$G(x_a, t_a; x_b, t_b)
=\left(\frac{m\omega_0}{2\pi \sinh \omega_0 T}\right)^{1/2}$$ $$
\exp \left\{ -\frac{m \omega_0}{2\sinh \omega_0 T} [(x^2_a+x_b^2)
\cosh \omega_0 T - 2 x_a x_b]\right\}$$ where $T=t_a-t_b$.
Changing $m\omega^2_0\to \frac{\omega^2}{4}, m=\frac12,$ and
identifying $x_a=x_b=0$, one arrives at the  result written in Eq.
(20).

\vspace{1cm}

{\bf Appendix 2}\\

{\bf The gluon  propagator in the einbein path-integral
representation}\\

 \setcounter{equation}{0} \def\theequation{A2.\arabic{equation}}

One starts with FSR for the free propagator which can be written
as
\be
G(x,y) = \int^\infty_0 ds (Dz)_{xy} \exp (-K)\label{A1.1}\ee and
introduce the einbein variable--dynamical mass $\mu(t)$ as  in
[10], so that $K$ can be rewritten as $$ K=m^2s+\frac14 \int^s_0
\left( \frac{dz_\mu(\tau)}{d\tau}\right)^2d\tau= $$
\be
= \int^T_0 dt \left\{ \frac{m^2}{2\mu(t)}+\frac{\mu(t)}{2} +\frac{
\mu(t)}{2} \left( \frac{dz_i(t)}{dt} \right)^2 \right\}.
\label{A1.2}\ee

 In $(Dz)_{xy},$
there is  integration over time components of the path, namely \be
(Dz_4)\equiv \prod_n \frac{d\Delta
z_4(n)}{(4\pi\varepsilon)^{1/2}} \delta \left(\sum\Delta
 z_4-T\right)\label{A1.3}\ee
 where $T\equiv x_4-y_4$, and using  the definition of $\mu(t), \mu(t) =\frac12 \frac{dt}{d\tau}$
  one can  rewrite the
 integration element in (\ref{A1.3}) as follows $(t\equiv z_4)$
 \be
 \frac{d\Delta z_4(n)}{\sqrt{4\pi \varepsilon}} = 2d\mu(n)
 \sqrt{\frac{\varepsilon}{\pi}} = \frac{d\mu(n) \sqrt{\Delta t}}{\sqrt{2\pi\mu(n)}},~~ \sqrt{\varepsilon}=
  \sqrt{\frac{\Delta
 t}{2\mu(n)}}.\label{A1.4}\ee
Moreover the $\delta$-function acquires the form
\be
\delta  \left(\sum \Delta z_4-T\right ) = \delta (s 2\bar
\mu-T)\label{A1.5}\ee where we have defined
\be
\bar \mu=\frac{1}{s} \int^s_0 2\mu(\tau) d\tau.\label{A1.6}\ee As
a result one can integrate in (\ref{A1.1}) over $ds$  using
$\delta$- function (\ref{A1.5}), and rewrite $ds (Dz)_{xy}$ as
$ds(D^4 z)_{xy}= (D^3 z)_{\vex \vey} D\mu$. One can write the
Green's function as follows \be G(x,y) = \int \prod
\frac{d^3\Delta z_i(n)}{l^3} e^{-K} \frac{d\mu (n)}{l_\mu(n)}
\frac{d^3p}{(2\pi)^3} e^{i\vep(\vex-\vey-\sum\Delta
\vez(n))}\label{A1.7}\ee where $K$ is given in (\ref{A1.2}), and
$l, l_\mu$  are $$ l(n) =\left( \frac{2\pi \Delta t}{\mu
(n)}\right)^{1/2},~~ l_\mu (n) = \left(\frac{2\pi \mu (n)}{\Delta
t}\right)^{1/2}, ~~ N\Delta t = x_4 - y_4 \equiv T.$$ The
integration over $d^3\Delta z_i(n)$ yields
\be
G(x,y) =  \int \frac{d^3p}{(2\pi)^3} e^{i\vep(\vex-\vey) -\frac12
\int^T_0 dt \mu(t)  \left( 1 + \frac{\vep^2+m^2}{\mu^2(t)}
\right)}\frac{1}{2\bar \mu }(D\mu).\label{A1.8}\ee

Taking into account the integral
\be
\int^\infty_0 \frac{d\mu(n)}{\sqrt{\mu(n)}}e^{-\frac{\Delta t}{2}
\left(\mu(n) + \frac{\vep^2+m^2}{\mu(n)}\right) }
=\sqrt{\frac{2\pi}{\Delta t}} e^{-\Delta t
\sqrt{\vep^2+m^2}}\label{A1.9}\ee one has  finally \be G(x,y)
=\int \frac{d^3p}{(2\pi)^3} \frac{e^{i\vep (\vex-\vey) - \int^T_0
dt \sqrt{\vep^2+m^2}}}{2\sqrt{\vep^2 +m^2}}\label{A1.10}\ee where
we have used relation following from  the stationary point in the
integral (\ref{A1.9})
\be
\bar \mu = \int^s_0 \mu(\tau) d\tau
=\sqrt{\vep^2+m^2}.\label{A1.11}\ee

This can be compared with the integral \be G(\ver, T) =\int
\frac{d^4p}{(2\pi)^4} \frac{e^{i\vep \ver + ip_4 T}}{p^2_4+\vep^2
+m^2}, ~~ \ver = \vex-\vey,\label{A1.12}\ee which reduces to
(\ref{A1.10}) after integrating over $dp_4$ for $T>0$. Eq.
(\ref{A1.12}) is the standard form of the free propagator. Now in
the case of interacting gluon as in section 4 one can still use
(\ref{A1.5}), (\ref{A1.6}) with the result, that $\bar \mu=\mu_0$,
with $\mu_0$ defined  in   (34) since $\mu_0$ does not depend on
time.


\begin{thebibliography}{99}
\bibitem{1}
 B.S.De Witt, Phys. Rev. {\bf 162}, 1195, 1239 (1967)\\
 J.Honerkamp, Nucl. Phys. B {\bf 48}, 269 (1972);\\
 G.'t Hooft Nucl. Phys. B {\bf 62}, 444 (1973), {\it  Lectures
at Karpacz, in : Acta Univ. Wratislaviensis }{\bf 368}, 345
(1976);\\ L.F.Abbot, Nucl. Phys. B {\bf 185}, 189 (1981).

 \bibitem{2} Yu.A.Simonov, Phys.  At.  Nucl.
  {\bf 58}, 107 (1995); Yad. Fiz. {\bf 58}, 113 (1995).

   \bibitem{3}
   Yu.A.Simonov, in: {\it Lecture Notes in Physics, v.479,
 p.144, Springer, 1996.}

\bibitem{4}
A.B.Kaidalov and Yu.A.Simonov, Phys. Lett. B {\bf 477} 163 (2000);
hep-ph/9912434;\\
       A.B.Kaidalov and
     Yu.A.Simonov,  Phys. Atom. Nucl. {\bf 63},  1428 (2000); Yad. Fiz. {\bf 63}  1428 (2000); hep-ph/9911291

 \bibitem{5}Yu.A.Simonov, Nucl. Phys. B (proc. Suppl) {\bf 23},
 283 (1991);\\
 Yu.S.Kalashnikova, Yu.B.Yufryakov, Phys. Lett.B {\bf 359}, 175
 (1995).

 \bibitem{6} Yu.S.Kalashnikova, D.S.Kuzmenko, Phys. Atom. Nucl.
 {\bf 64}, 1716 (2001);  ibid. {\bf 66}, 955 (2003),
 hep-ph/0203128; hep-ph/03022070.


\bibitem{7} Yu.A.Simonov, Nucl. Phys. B {\bf 592},  350 (2001);
hep-ph/0003114.


\bibitem{8}
A.Di Giacomo and H.Panagopoulos, Phys. Lett. B {\bf 285}, 133
(1992);\\
 A.Di Giacomo, E.Meggiolaro and  H.Panagopoulos, Nucl. Phys.
 B {\bf 483}, 371  (1997);\\ M.D'Elia, A.Di Giacomo and E.Meggiolaro
Phys.Lett.  B {\bf 408}, 315 (1997).

\bibitem{9} K.J.Juge, J.Kuti and C.Morningstar, Nucl. Phys. Proc.
Suppl. {\bf 73}, 360 (1999); hep-lat/9809015; Phys. Rev. Lett.
{\bf 90}, 161601 (2003).

\bibitem{10}
A.Yu.Dubin, A.B.Kaidalov, and Yu.A.Simonov, Phys. Lett. B {\bf
323},  41 (1994);\\
 A.Yu.Dubin, A.B.Kaidalov, Yu.A.Simonov, Phys. At. Nucl.  {\bf 56}, 1745 (1993);
 hep-ph/9311344.


\bibitem{11} Yu.A.Simonov and J.A.Tjon, Ann Phys. {\bf 228}, 1
(1993);\\
 Yu.A.Simonov, J.A. Tjon, Annals Phys.  {\bf 300}, 54 (2002);
hep-ph/0205165.

\bibitem{12} Yu.A.Simonov, Phys. Rept. {\bf 320}, 265 (1999); JEPT
Lett {\bf 69}, 505 (1999).

 \bibitem {13}
  H.G.Dosch, Phys. Lett. B {\bf 190}, 177 (1987);\\
  H.G.Dosch and Yu.A.Simonov, Phys. Lett. B {\bf 205}, 339 (1988);\\
   Yu.A.Simonov, Nucl.  Phys.  B {\bf 307}, 512 (1988).


  \bibitem{14}   G.Bali, Nucl. Phys. Proc. Suppl.  {\bf 83}, 422
  (2000), Phys. Rev.  D{\bf 62}, 114503 (2000);\\ A.Hasenfratz, R.Hoffmann and
 F.Knechtli, Nucl. Phys. Proc. Suppl. {\bf 106}, 418 (2002).


\bibitem{15} B.Bolder, T.Struckmann, G.S.Bali, N.Eicker et al.,
Phys. Rev. {\bf D63}, 074504 (2001).

\bibitem{17}
M.L\"{u}scher and P.Weisz, JHEP 0207, 049 (2002).

 \bibitem{17} E.Eichten et al. Phys. Rev. D {\bf 21}, 203 (1980);\\
E.J.Eichten, K.Lane and C.Quigg, Phys. Rev. Lett. {\bf 89}, 162002
(2002).
\bibitem{18} A.M.Badalian, B.L.G.Bakker, Phys. Rev. D {\bf 67},
071901 (2003);\\
 A.M.Badalian, B.L.G.Bakker and A.I.Veselov, Phys. Rev. D  (in press),
 hep-ph/0311010.

\bibitem{16} R.P.Feynmann, A.R.Hibbs, {\it  Quantum Mechanics and Path
Integrals Mc.Graw-Hill, N.Y., 1965, problem 3.8.}

 \end{thebibliography}
\end{document}